\documentclass{JHEP3}
\usepackage{graphicx}

\def\hbar{\hspace{0pt}\raisebox{1pt}{$-$} \hspace{-7pt} h}

\def\5{\overline 5}

\newcommand{\ba}{\begin{eqnarray}}
\newcommand{\ea}{\end{eqnarray}}
\newcommand{\no}{\nonumber}
\newcommand{\be}{\begin{equation}}
\newcommand{\ee}{\end{equation}}
\newcommand{\bea}{\begin{eqnarray}}
\newcommand{\eea}{\end{eqnarray}}

\def\NG{Nambu-Goldstone }

\def\SM{standard model~}
\def\RS{Randall-Sundrum~}
\def\EWSB{electro-weak symmetry breaking~}
\def\EW{electro-weak~}
\title{Phenomenology of a light scalar:
the dilaton}
\date{\today
}
\author{Luca~Vecchi\footnote{vecchi@lanl.gov}\\
Theoretical Division T-2, Los Alamos National Laboratory\\
  Los Alamos, NM 87545, USA}
\abstract{
We make use of the language of non-linear realizations to analyze
\EWSB scenarios in which a light dilaton emerges from the breaking
of a nearly conformal strong dynamics, and compare the phenomenology of the dilaton to that of the well motivated light
composite Higgs scenario. We argue that -- in addition to departures in the decay/production rates
into massless gauge bosons mediated by the conformal anomaly -- characterizing features of the light
dilaton scenario (as well as other scenarios admitting a light CP-even scalar not directly related to the breaking of the electro-weak symmetry) are off-shell events at high invariant mass involving two longitudinally polarized vector bosons and a dilaton, and tree-level flavor violating processes. Accommodating
both \EW precision measurements and flavor constraints appears
especially challenging in the ambiguous scenario in which the
Higgs and the dilaton fields strongly mix. We show that warped
higgsless models of \EWSB are explicit and tractable realizations
of this limiting case.

The relation between the naive radion profile often adopted in the study of holographic realizations of the light dilaton scenario and the actual dynamical dilaton field is clarified in the Appendix.

}
\preprint{LA-UR 09-06916}
\maketitle
\begin{document}
\section{Introduction}

The detection of the Higgs boson is one of the most ambitious aims
of the CERN large hadron collider (LHC). Precision measurements
strongly suggest the presence of this excitation in the hundred
GeV range, but its very existence seems to be in contrast with our
understanding of naturalness. Any alternative \EWSB (EWSB) sector
has to explain this tension.

If the forthcoming experiments detect a light CP-even scalar but
no accompanying states -- the latter being too heavy or too broad to
be directly observed -- it would turn out to be crucial to understand whether the light scalar \textit{is} the actual Higgs, namely the physical component of an electroweak doublet. In
this letter we would like to address this issue using a
phenomenological approach.

A parametric separation between the mass of the light scalar and
those of the heavy states is most naturally justified by a
strongly coupled EWSB theory in which the scalar emerges as an
approximate Goldstone boson of some broken global symmetries. We
focus on two classes of models which are consistent with our
assumptions. The first class is a deformation of the \SM Higgs sector, in which the CP-even scalar emerges as the fourth
excitation of a light composite Higgs doublet. The physics of this strongly coupled light Higgs
(SILH) has been studied in detail in~\cite{SILH}.

The second class of models include scenarios in which the strong
EWSB sector is a nearly conformal field theory (CFT) that admits a
light dilaton in the spectrum~\cite{GGS}. In this class the
CP-even scalar emerges as an approximate Goldstone boson of the
breaking of the conformal symmetry down to the Poincar\'e group,
and it is not directly related to the EWSB sector. Appropriate
field theories are expected to reproduce the invoked mechanism in
a quantum mechanical context. However, a generic strong dynamics is not a
realistic candidate. An heuristic motivation is that in the latter
case the conformal breaking is governed by the running of the gauge
coupling; in the IR the coupling is strong and there is no small
parameter suppressing the dilaton mass compared to the dynamically
generated scale~\cite{RZ}. Lattice simulations confirm these conclusions.

This picture changes if the small
coupling is taken to be $1/N$, where $N$ is the number of
fundamental constituents. Evidences in favor of this mainly come from the gauge/gravity correspondence (see~\cite{Piai} for a recent example). The Randall-Sundrum model~\cite{RS}, as well as
several other models for physics beyond the \SM belonging to the
same universality class, are well known realizations of the light dilaton scenario. Supersymmetry then offers a simple way to relate the presence of a light $\eta'$ and the dilaton\footnote{I am indebted to Sergio
Cecotti for an illuminating discussion of this point.}. Indeed, in
a supersymmetric theory the scale current and the axial current are part of the same multiplet. If there exists a limit (the large N limit in a nonsupersymmetric theory!) in which the anomalous axial symmetry is exact, then in a non-chiral vacuum an approximate Goldstone boson (GB), the so called
$\eta'$, would emerge along with a CP-even pseudo-GB partner, i.e. the
dilaton. All of these considerations suggest that theories with many fundamental degrees of freedom may represent the natural framework for the realization of the light dilaton scenario. This would have implications in the study of the phenomenological signatures of this class of models.

It is very important to appreciate that, in the case the operators responsible for EWSB and the spontaneous breaking of the CFT do coincide, the Higgs and the dilaton are the very same field. This is what
happens at the classical level in the \SM~\cite{Ellis}, and what
is expected to occur in well calibrated walking technicolor
scenarios~\cite{Sannino}. In this limit the SILH and the light dilaton effective theories overlap. This intrinsic ambiguity suggests that a phenomenological dicrimination between the two scenarios is a highly non-trivial task, in general. Nevertheless, we believe that the identification of features that most naturally characterize one theory as opposed to the other, and in particular a study of how natural is the above limit and what the phenomenogical implications are, has physical relevance. This is one of the aims of the present study.

\section{Phenomenology of a light scalar}

A light and chargeless CP-even scalar can arise as the physical
excitation of a Higgs doublet, but in principle it may also arise
as a generic pseudo Goldstone boson of an appropriately engineered
strong dynamics. In the latter case the scalar field would have no
direct relations with the Higgs sector. In this paper we address the question whether these two classes of
scenarios can be disentangled under the hypothesis that the only
new physics detected in the forthcoming experiments is the light
CP-even scalar. To be definite we choose a representative of each
class: from the first class we consider the strongly interacting
light Higgs scenario (SILH), while from the second we consider the
strongly interacting light dilaton (SILD) scenario.

Following the notation introduced in~\cite{SILH}, we decide to
describe our models by using three free parameters: $f$,
$g_\rho$, and $m$. The first parameter is the energy scale $f$ at
which the strong dynamics spontaneously breaks its approximate
global symmetry, and hence leading to the appearance of a
Goldstone doublet (SILH) or singlet (SILD). The broken symmetry
can be either a global one, as in the case of the SILH scenario,
or a space-time symmetry, as in the case of the dilaton scenario.
The scale $f$ is basically a measure of the strength of the Goldstone
boson interactions. The second parameter, $g_\rho$, is the typical coupling
of the massive resonances, which in particular is
expected to enter into the definition of the masses of the heavy
composites as a proportionality factor, $m_\rho\propto g_\rho$.
The statement that the new physics is strong compared to the \SM
can be expressed via the relation $g_{SM}\ll g_\rho$, where $g_{SM}$
is a typical \SM coupling and $g_\rho=4\pi$ for a maximally strong
dynamics. The third parameter is the mass of the pseudo Goldstone
boson, $m$, and crucially depends on the explicit source of
symmetry breaking. The rules of non-linear realizations are expected to hold up to corrections of order $m^2/f^2\ll1$.

The physics of both the SILH and SILD can be captured by the
effective lagrangian of a generic chargeless CP-even scalar $S$:
\ba\label{gen}
 {\cal L}_{eff}
&=&\frac{1}{2}m_V^2V_\mu^2\left(1+2a_1\frac{S}{v}+a_2^2\frac{S^2}{v^2}\right)\\\no
&+&\bar \psi_i\psi_j
\left[m_i\delta^{ij}\left(1+b\frac{S}{v}\right)+b^{ij}\frac{S}{v}\right]\\\no
&+&c\frac{g_{SM}^2}{16\pi^2}F_{\mu\nu}^2\frac{S}{v}+\dots \ea
We allowed the presence of a higher
dimensional operator $SF_{\mu\nu}^2$, where $F_{\mu\nu}$ is the
photon or gluon field strength, because no such a coupling is
present at leading order, and it may play a phenomenological role
if the scale of new physics is not too high~\cite{MW}. 

The \SM Higgs boson is just a particular case
of the theory~(\ref{gen}) with $a_1=a_2=b=1$ and $b_{ij}=c=0$. For
compliteness we mention that the integration of the top induces a
coupling $c=O(1)$.

\paragraph{The SILH:}

In the SILH model the full Higgs doublet emerges as a Goldstone
field, the massive composites being naturally at the scale
$m_\rho= g_\rho f$. The explicit violation of the global symmetry
induces the generation of a potential for the doublet and
eventually implies EWSB. In the broken \EW phase the Higgs boson
is the only physical and light relic of the strong dynamics. A
generic coupling of the SILH to the \SM vectors and fermions
differs by an $O(v^2/f^2)$ correction with respect to the \SM Higgs, and
approaches the latter as the heavy states decouple, $v\ll f$.
Although this regime appears unnatural, it turns out that a mild
suppression $v<f$ is sufficient for fitting the \EW precision
data. In this model the authors of~\cite{SILH} found that the physics of the light scalar can be effectively described by the lagrangian~(\ref{gen}) with:
\ba\label{SILH} a_1=1-c_H\frac{v^2}{2f^2};\quad
a_2^2=1-2c_H\frac{v^2}{f^2};\quad
b=1-(c_y+c_H/2)\frac{v^2}{f^2};\quad b_{ij}=0;\quad c=0, \ea where
$c_{y,H}$ are $O(1)$ parameters defined in~\cite{SILH}. In the
rest of the paper we will be mainly concerned with the dilaton
physics. We refer the reader to the paper~\cite{SILH} for more
details on the SILH scenario.

\paragraph{The SILD:}

In the SILD the only Goldstone mode emerging at a scale $f$ is the
dilaton itself, while EWSB is triggered by a strong dynamics at a
somewhat lower scale $v=250$ GeV. For practical purposes the \SM
symmetry can be taken to be non-linearly realized below the
characteristic mass scale of the strong Higgs sector,
$m_\rho=g_\rho v$. Notice that the light scalar $S$ in this case
is not the physical excitation of a Higgs doublet. The couplings
of the dilaton to the \SM fields can be derived by applying the
theory of phenomenological lagrangians. For completeness we now
briefly review the main aspects of the procedure, for a detailed
derivation see for instance~\cite{Salam}.

The conformal symmetry is an extension of the Poincar\'e algebra
which includes scale invariance as well as the so called special
conformal transformations. By definition, a CFT is necessarily
scale invariant, but the converse need not be true. Let us first
focus on the implications of scale invariance on the effective
field theory. Given a local operator ${\cal O}(x)$ of scaling
dimension $\Delta$, the transformation under dilatations
$x^\mu\rightarrow e^\lambda x^\mu$, where $\lambda$ is an
arbitrary constant, is given by ${\cal O}(x)\rightarrow
e^{\lambda\Delta}{\cal O}(e^\lambda x)$. The theory is invariant
under scaling if the lagrangian has dimension $\Delta=4$. For $\Delta\neq4$ the scale symmetry can be locally
restored by introducing an appropriate GB field $\sigma$, the
dilaton. This field transforms in-homogeneously under a dilatation
$\sigma\rightarrow\sigma+f\lambda$ and thus signals the
spontaneous breakdown of the $\sigma=0$ vacuum. In complete
analogy with the non-linear realization of internal symmetries, it
is convenient to introduce a field transforming linearly under
scale transformations: \ba\label{dilaton} \chi(x)\equiv
fe^{\sigma(x)/f}\rightarrow e^\lambda\chi(e^\lambda x) \ea For
practical purposes we can identify
$\bar\chi=\chi-f=\sigma+O(\sigma^2)$ as our dynamical field rather
than $\sigma$ itself. Because $\sigma$ and $\bar\chi$ generate the same one-particle states, the
S-matrices of the two fields do coincide. A dimension-4 operator can finally be obtained by
replacing the dimension $\Delta$ operator ${\cal O}$ with ${\cal
O}(\chi/f)^{4-\Delta}$.

At the classical level this is basically what the \SM Higgs does~\cite{Ellis},\cite{GGS}.
To appreciate this, let us derive the leading couplings of
the dilaton to the spin-1, spin-1/2, and the Goldstone $SU(2)$
matrix $U$ of the \SM $SU(2)\times U(1)\rightarrow U(1)$ symmetry
breaking pattern by assuming that the \SM fields have a classical scaling: 
\ba\label{embed}
 {\cal L}
&=&\frac{v^2}{4}Tr|D_\mu U|^2\left(\frac{\chi}{f}\right)^2+m_i\bar
\psi_i
U\psi_i\left(\frac{\chi}{f}\right)+\dots\\\no
&=&
\frac{1}{2}m_V^2V_\mu^2\left(1+\frac{\bar\chi}{f}\right)^2+m_i\bar
\psi_i
\psi_i\left(1+\frac{\bar\chi}{f}\right)+\dots
\ea 
The model~(\ref{embed}) corresponds to the case 
\ba\label{SILD}
a_1=a_2=\frac{v}{f};\quad b=\frac{v}{f};\quad b_{ij}=0 
\ea 
and proves that, apart from a $v/f$ rescaling, the couplings of the dilaton are formally those of a fundamental Higgs.

The $\Delta>4$ operators are either irrelevant for our
purposes or severely constrained by \EW precision data\footnote{In this
regard we should mention that a light dilaton is expected to
alleviate the tension between the scale $m_\rho$ of new physics
and the \EW precision measurements by screening the large
corrections of the strong dynamics to the precision
parameters. The effect is estimated to be small -- see the quantitative discussion in~\cite{Barbieri} -- so that the higher
dimensional operators can be neglected in our discussion. Yet, it would be interesting to understand the
relation between this screening effect and the slight improvement
on the \EW $S$-parameter fit that a walking dynamics, as opposed
to a generic strongly coupled theory, is expected to cause.}.
An exception appears to be the coupling to the unbroken gauge
bosons. The latter arises as an effect of the scale-dependence of
the renormalized coupling $g(\mu)$, i.e. the so called scale
anomaly. The prescription for the non-linear realization of the
scale symmetry illustrated above may be equivalently restated by
saying that the renormalization scale $\mu$ must be compensated
by an appropriate power of the dilaton field
$\mu\rightarrow\mu\chi/f$. Hence, if we assume that the gauge symmetry of the \SM is part of the CFT, the coupling
\ba
-\frac{1}{4g_{SM}^2}F_{\mu\nu}^2=+\frac{\beta_{SM}}{2g_{SM}}F_{\mu\nu}^2\frac{\bar\chi}{f}+\dots
\ea
should be added to~(\ref{embed}). (Notice that on the right hand side of the equation the vectors have been canonically
normalized). This is equivalent to the introduction of a coefficient:
\ba\label{c}
c=\frac{8\pi^2\beta_{SM}}{g_{SM}^3}\frac{v}{f}
\ea
in~(\ref{gen}). The
phenomenology of~(\ref{SILD}) plus~(\ref{c})  has been studied by the authors
of~\cite{GGS}. Because the coefficient~(\ref{c}) can potentially be larger than the top loop contribution, the anomalous term may well represent the most accessible
signal of the SILD, especially in the ambiguous regime $v\sim f$~\cite{GGS}.

If the \SM gauge symmetry represents an explicit breaking of the scale invariance, then the dilaton couplings to the unbroken gauge group is no more predicted by the trace anomaly, and $c$ is no more given by~(\ref{c}).
The latter coupling is more generally mediated by mixing between
composites and \SM vectors, and by the
integration of charged heavy composites. The resulting operator
has been included in~(\ref{gen}) with a model dependent $c=O(v/f)$
factor. Such a contribution can be in principle sizable if the
scale invariant theory has a large number of degrees of freedom at
the scale $g_{\rho}f$, and may compete with the top loop irrespective of whether the gauge symmetry is or is not embedded into the broken CFT.

In the next few sections we will also argue that the lagrangian~(\ref{embed}) is not accurate if the Higgs sector has non-negligible anomalous dimensions, as it is expected from a strongly coupled sector, or if the CFT flavor structure is non-trivial, as it is expected in a generic model addressing the \SM fermions hierarchy. In particular we will see that the relations $a_1^2=a_2^2$ and $b^{ij}=0$ are not characteristic features of the light dilaton scenario.

We conclude this section by mentioning that a phenomenologically acceptable realization of both the SILH and
SILD scenarios requires the existence of an explicit breaking of
the global symmetries of the strong dynamics, i.e. the generation
of a mass $m$ for the light scalar. The source of explicit
breaking is typically the \SM in the SILH scenario~\cite{SILH},
while, as already stressed in the introduction, it may be the \SM as well as the strong dynamics itself in the SILD
model. A deformation of the CFT generally implies corrections to
the dilaton vacuum expectation value with respect to the symmetric
solution. The true vacuum can thus differ from $f$, causing a shift in the dilaton couplings. The
relation $f<v$ may be in principle allowed in this case, but the validity of our phenomenological approach would come into question. We will comment more on the magnitude of the ratio $v/f$, and its physical meaning, in a following section. For the moment we emphasize that the phenomenology of the SILD approaches that of a SILH in the regime
$v\sim f$, whereas that of an
ordinary Higgsless theory for $v/f\ll1$.

\paragraph{Implications of the special conformal transformations:}

The additional constraints coming from the special conformal transformations affect the derivative
couplings only. Given a local operator ${\cal O}$ of weight
$\Delta$ and transforming under the Lorentz group as an
irreducible representation with infinitesimal generators
$S_{\mu\nu}$, we can define a conformally invariant derivative
as~\cite{Salam} \ba \label{der}D_\mu{\cal
O}=\left[\partial_\mu+\left(iS_{\mu\nu}-\Delta\eta_{\mu\nu}\right)\partial_\nu\log\chi
\right]{\cal O}.\ea An important point to be noticed is that the
non-linear realization of the full conformal group does not
require the introduction of additional Goldstone bosons: the
broken special conformal transformations do not generate new
physical excitations from the vacuum. Contenting ourselves with a
leading order expansion in external momenta, the derivative
interaction of the dilaton field required by~(\ref{der}) has
physical impact only if ${\cal O}$ is a scalar field with
non-vanishing vacuum expectation value (see section 5). Being
mainly interested in the dilaton couplings to spin $1$ and $1/2$, we conclude that the leading low energy effective field theory is not sensitive to whether the UV
completion possesses a full conformal invariance or just a scaling symmetry.

\paragraph{The radion:}

The model~(\ref{SILD}) plus~(\ref{c}) has a dual interpretation in terms of the
RS1 scenario~\cite{RS} with a heavy Higgs and the \SM fields
placed on the IR brane. Yet, as soon as the SM fields move away from the IR brane the relations~(\ref{SILD}) and~(\ref{c}) no longer hold.

The presence of the IR brane in the bulk
AdS background can be interpreted as a spontaneous breaking of the
conformal symmetry of the strongly coupled 4D dual theory, and the
gauge/gravity interpretation of the dilaton is given in terms of
the radion field~\cite{RZ}~\cite{AHPR}. The dynamical description
of the radion field is encoded in the perturbed line
element~\cite{CGR} \ba\label{rad}
ds^2&=&e^{-2ky+Qe^{2ky}}\eta_{\mu\nu}dx^\mu
dx^\nu-\left(1-Qe^{2ky}\right)^2dy^2\\\no
&=&e^{-2kw}\eta_{\mu\nu}dx^\mu dx^\nu-dw^2+\dots \ea where $Q(x)$ represents the 4D field appearing in
the effective field theory. The solution $Q=0$ is the conventional
Randall-Sundrum background.

 In the second line of~(\ref{rad}) we made a change of variable to the new coordinate $w(y,x)$,
 defined by $2kw=2ky-Qe^{2ky}$. In terms of it the metric can be expressed in a compact
form where the dynamical field $w$ determines the spacetime-dependent proper length of
the extra dimension with classical vacuum $y$. At the quantum level the extra dimension
can thus be thought of as a physical degree of freedom, the dilaton of
the dual theory. Notice in fact that the model preserves a
dilatation symmetry $x^\mu\rightarrow e^\lambda x^\mu$ at the
perturbed level if the field $w$ transforms in-homogeneously
$kw\rightarrow \lambda +kw$, in complete analogy with $\sigma$
in~(\ref{dilaton}).

The second line of~(\ref{rad}) has the same form as the naive
ansatz for the radion profile proposed originally by Randall and
Sundrum~\cite{RS}, up to $O\left(\partial_\mu w\right)$ terms having subleading impact on the
4D physics. This clarifies why the physics of the actual
dynamical mode $Q$ agrees (up to derivative couplings and hierarchically suppressed
corrections) with the results found for the naive ansatz (see for
instance~\cite{RZ},\cite{GW},\cite{CGK},\cite{Peloso}, and
references therein), although the latter is not a dynamical mode. See the Appendix for a discussion of this point.

\section{Light scalars and EWSB}

Suppose the LHC detects a light CP-even scalar and no other
states: can we tell whether this particle is really the physical
excitation of a Higgs doublet?

Thanks to
the equivalence theorem, at external momenta $p^2\gg m_V^2$ the
three Goldstone boson fields $\pi^a$ belonging to the unitary
matrix $U=e^{i\pi^a\sigma^a/v}$, where $\sigma^a$ are the Pauli
matrices, can be identified with the longitudinally polarized
vector bosons $V_L^a$. Thus, we can directly probe the strong dynamics by focusing on the physics of the scalars $\pi^a$ and $S$, and rewriting the first line of~(\ref{gen}) as\footnote{This section has some overlaps with the analysis presented in~\cite{contino} and~\cite{Grojean}.}
\ba \label{Spp}
\frac{v^2}{4}Tr\left(\partial_\mu
U^\dagger\partial^\mu
U\right)\left(1+2a_1\frac{S}{v}+a_2^2\frac{S^2}{v^2}\right).\ea
The lagrangian~(\ref{Spp}) should be supplemented with a potential
for the scalar $S$; however, being mainly interested in the high energy
behavior of the amplitudes, this term can be discarded.

Typical values of the parameters $a_1$ and $a_2$ for the SILH are given up to $O(v^2/f^2)$ in~(\ref{SILH}), and for
the SILD in~(\ref{SILD}), under the assumption of negligible anomalous dimensions. In both cases we see that the elastic scattering of
longitudinally polarized vectors violates perturbative unitarity at the scale
of compositeness of the Higgs sector:
\ba\label{pppp} {\cal A}(\pi^a\pi^a\rightarrow\pi^b\pi^b)=
\frac{s}{v^2}(1-a_1^2),\ea with $s$ the Mandelstam center of mass
energy and $a\neq b$. These processes can be effectively used to
discriminate between the two scenarios only if we are able to probe the heavy resonances, at scales of order $g_\rho v$ and $g_\rho f$ for the SILD and SILH respectively. Even if the heavy states are not detected, this channel will provide crucial informations regarding the strength of the Higgs sector, see~\cite{maina} for a recent study.

Another interesting channel is the elastic scattering $2S\rightarrow2\pi$~\cite{Rattazzi}, for which
\ba\label{SSpp} {\cal A}(SS\rightarrow\pi^a\pi^a)=
\frac{s}{v^2}(a_1^2-a_2^2).\ea If we neglect the anomalous dimensions of the Higgs sector of the CFT, $a_1^2=a_2^2$, and the latter amplitude vanishes for $S$ the dilaton, as in the \SM Higgs
scenario; the observation of high energy $2S\rightarrow2\pi$ events in this limit would be a characterizing feature of the SILH. More generally, if we allow for non-zero anomalous dimensions of the CFT Higgs sector (we will see in section 5 that this implies $a_1^2\neq a_2^2$), then~(\ref{SSpp}) represents an indisputable evidence in favor of the compositeness of the
EWSB sector.

If we were able to isolate the strong dynamics from the explicit
symmetry breaking, though, the Goldstone boson sectors of the SILH and SILD scenarios would
look radically different. On the one hand, the SILH
scenario, as the fundamental Higgs of the standard model,
possesses an $O(4)$ symmetry rotating the 4 scalars $\pi^a,S$. At scales $f^2\gg p^2\gg m^2$ the Goldstones $\pi$
effectively recombine with the SILH in the full order
parameter, i.e. the Higgs doublet, and the $O(4)$ symmetry manifests itself in relations between the $S,\pi$ scattering amplitudes\footnote{The $O(v^2/f^2)$ corrections are next to leading order in the expansion adopted in~\cite{SILH}, but are already present at the 2-derivative level.}
\ba\label{O}
{\cal A}(SS\rightarrow2\pi^a)={\cal A}(2\pi^a\rightarrow2\pi^b)\left(1+O\left(\frac{v^2}{f^2}\right)\right),
\ea
(compare~(\ref{SSpp}) with~(\ref{pppp}) by making use of~(\ref{SILH})), and in the suppression of $O(4)$ violating events as $S\rightarrow2\pi$. On the other hand, no
such symmetry is present in the SILD model or any model in which the light scalar $S$ is not part of the weak doublet. In the latter case, relations of the type~(\ref{O}) are not observed, whereas events like $\chi\rightarrow2\pi$ are allowed by the strong dynamics and are therefore enhanced at high energies.

These observations reveal that a discrimination between the two models may be feasible in the limit in which the corrections in~(\ref{O}) are sufficiently small.
For example, if the SILH had, say, a $v^2/f^2$ of the order of a few percent, while the SILD $v/f\sim1$ at the ten percent level, then the two scalars would look very much like the fundamental Higgs boson for what concerns the couplings to the \SM fields, but they would present a radically different UV physics for what concerns the couplings to the unphysical Goldstone bosons. First, the SILH would satisfy the crucial relation~(\ref{O}), where the amplitudes are enhanced at high energy ${\cal A}\sim s/f^2$ as opposed to a fundamental Higgs. This would lead to interesting observable consequences, like the sum rule eq.(4.19) in~\cite{SILH}. Second, observations of energy enhanced off-shell $O(4)$
violating processes $S\rightarrow 2V_L$ (in vector boson fusion as well as associated production events) would be a distinctive signature of the
dilaton model. These processes may be for example tested as excesses in $t\bar t\rightarrow
\chi\rightarrow V_LV_L'(\rightarrow 4l,ll\nu\nu)$ events at large
invariant mass. 

The deviations in the hard process $t\bar t\rightarrow
S\rightarrow V_LV_L$ are
potentially significant, and reach the $100\%$ at moderately
high energy scales $\sqrt{s}>3m$ even for $O(10\%)$ departures from the standard model couplings. Given such a sensitivity in the hard process, the potentialities of these events in the extraction of the relevant Higgs couplings are remarkable. At the LHC, even after a relatively low luminosity $\sim30$ fb$^{-1}$, it would be possible to achieve accuracies as high as $5-10\%$, thus approaching the noise of a $\sim2\sigma$ deviation~\cite{Zepp}.

Processes involving only the light scalar $S$ are generally more
difficult to observe but are also useful to probe the strong
dynamics. At low energies the scalar self-couplings are dominated by the
potential interactions, and the extrapolations of trilinear ($g_{SSS}$) and quartic ($g_{SSSS}$) couplings would provide useful informations on the source of explicit breaking of the Goldstone symmetry~\cite{SILH},~\cite{GGS}. Large $O(1)$ deviations from the \SM Higgs trilinear
coupling may in principle be observed at the LHC in a limited mass range
and after an integrated luminosity of about $300$
fb$^{-1}$ is attained. An
international linear collider (ILC) would certainly be more adequate, and may be able to test the triple coupling
with a much larger accuracy, of order some $10{\%}$.

At sufficiently large energies these measurements are, however, substantially altered by the strong dynamics.
In the absence of an explicit breaking of the global invariance the scalar self-couplings are derivatively induced. The SILH self-couplings are governed in the high energy regime by the $O(4)$ invariance and, for instance: 
\ba\label{SSS} 
{\cal A}_{SILH}(2S\rightarrow2S)=
c_H\frac{s}{f^2}.
\ea 
The SILD lagrangian is, up to $O(p^4)$, 
\ba\label{self}
\frac{(\partial_\mu\chi)^2}{2}+\frac{c_3}{(4\pi)^2}\frac{\partial_\mu\chi\partial_\nu\chi\partial^\mu\partial^\nu\chi}
{\chi^3}+\frac{c_4}{(4\pi)^2}\frac{(\partial_\mu\chi\partial_\nu\chi)^2}{\chi^4}+\dots,
\ea where for brevity we neglected operators that give vanishing contribution on shell. The coefficients $c_i$ can be estimated using naive dimensional analysis\footnote{As opposed to the chiral lagrangian the interactions start at $O(p^4)$ -- although this property is a bit obscured in terms of the variable $\sigma=f\log\chi/f$. In a cutoff regularization scheme, we can estimate the coefficients by requiring for example that the operators $c_i$ ($i=3,4$) induce small corrections on the kinetic term (this is not a relevant correction on shell, but our conclusions are completely general). This condition requires $c_i<(4\pi/g_\rho)^i$, where we cutoff the loop integrals at the physical scale $g_\rho f$. For $g_\rho\sim4\pi$ our estimate $c_i=O(1)$ agrees with the rules proposed in~\cite{SILH}.}, and are expected to be of order unity in a strong dynamics. Both $2\chi\rightarrow2\chi$ and the ("$O(4)$ violating") process $2\chi\rightarrow\chi$ start at $O(p^4)$ and are suppressed in the regime of validity of
our effective description. 

In both the SILH and the SILD, the strong dynamics modifies significantly the scalar
vertices at energies of the order of $f$ or bigger. This observation is potentially relevant
for the SILD physics at a linear collider, especially in the regime
$v\sim f$, as the $O(p^4)$ terms would come into play at the scales ($500-1000$ GeV) at which the ILC operates. As an instance, for $v/f=1+10\%$ a perturbative description of the $V_LV_L$ scattering may be delayed up to $\sim4$ TeV, whereas a strong $\chi$ self-coupling would enter around $1$ TeV.

\section{Dilaton couplings to \SM fermions}

Tree level flavor violations mediated by a composite Higgs are typically suppressed if the Higgs
is a SILH~\cite{Contino}. We now show that this conclusion does not apply to many realistic realizations of the SILD
scenario.

The dilaton couplings
to the \SM fermions are expected not to mediate flavor changing
neutral currents (FCNC) only if the \SM fermions couple to CFT operators with flavor universal scaling dimensions. This is
certainly realized if the \SM fermions have all the same scaling
representation. If this is the case, the most general
Yukawa coupling for the dilaton can be written after EWSB as 
\ba\label{FC}
F_{ij}v\,\bar\psi_i\psi_j\left(\frac{\chi}{f}\right)^{\Delta}=m_{ij}\bar\psi_i\psi_j\left(1+\Delta\frac{\bar\chi}{f}+\dots\right)
\ea 
where $\Delta$ is an arbitrary number, $i,j$ are flavor
indeces, and $F_{ij}$ is a dimensionless scale invariant function,
which we can identify with the Yukawa matrix. If the CFT operators do not have a flavor universal scaling, or if
the CFT is broken, in the low energy dynamics it would emerge a
non-scale invariant function of the dilaton
$F_{ij}=F^{(0)}_{ij}+F^{(1)}_{ij}\bar\chi/f+\dots$, with
$F^{(a)}_{ij}$'s naturally of the same order. Equivalently, the
dimension $\Delta$ in~(\ref{FC}) would become flavor-dependent and would generally lead to flavor violating events.

It is instructive to show how these conclusions arise by looking at
specific UV realizations. One can identify two distinct ways to
generate the \SM flavor structure. The first is the one
implemented in composite Higgs models and (extended) technicolor
models, and contains the minimal flavor violation (MFV) class.
This is based on the coupling \ba\label{class1} {\cal
L}_{1}=y_{ij,a}\bar\psi_i\psi_j{\cal O}_{a}, \ea where the CFT
operators ${\cal O}_{a}$ have generally different scaling
dimensions $\Delta_{a}$. After EWSB the low energy effective
theory is obtained by integrating out the energetic fluctuations
of the CFT and reads \ba\label{1} {\cal
L}_{1}&=&y_{ij,a}\,\bar\psi_i\psi_j \langle{\cal
O}_{a}\rangle\left(\frac{\chi}{f}\right)^{\Delta_{a}}+\frac{y^2}{m_\rho^2}\bar\psi\psi\bar\psi\psi+\dots\\\no
&=&m_{ij}\bar\psi_i\psi_j\left(1+\Delta_{ij}\frac{\bar\chi}{f}+\dots\right)+\frac{y^2}{m_\rho^2}\bar\psi\psi\bar\psi\psi+\dots
\ea The coefficients of the four fermion operator is a symbolic
representation for the sum $y^2/m_\rho^2= y_{ij,a}D_{ab}y_{kl,b}$,
where the $D$ matrix originates from the propagator of the
operators ${\cal O}_{a}$. It is important to emphasize that the
FCNC couplings are there even if the \SM fermions are part of the
CFT, for example if $\Delta_a=4-\Delta_i-\Delta_j$. For a
universal dimension $\Delta_{a}=\Delta$ we recover~(\ref{FC}),
that is $\Delta_{ij}=\Delta$.

The second class approaches the flavor problem by employing a
seesaw type mechanism. This is the class usually implemented in
the Randall-Sundrum scenarios with \SM fermions in the
bulk~\cite{PG}, see~\cite{CP} for the dual interpretation. The
bare lagrangian now reads: \ba\label{class2} {\cal
L}_{2}=\lambda^{ia}_L\psi^i_L{\cal
O}^a_R+\lambda^{jb}_R\psi^j_R{\cal O}^b_L \ea and leads to an
effective field theory of the form \ba\label{2} {\cal
L}_{2}&=&\lambda^{ia}_L\lambda^{jb}_R\langle
D^{ab}\rangle\psi_L^i\psi_R^j\left(\frac{\chi}{f}\right)^{\Delta_R^a+\Delta_L^b-4}+
\frac{\lambda^4}{m_\rho^2}\bar\psi\psi\bar\psi\psi+\dots\\\no
&=&\lambda^{ia}_L\lambda^{jb}_R\langle
D^{ab}\rangle\psi_L^i\psi_R^j
\left(1+(\Delta_R^a+\Delta_L^b-4)\frac{\bar\chi}{f}+\dots\right)+
\frac{\lambda^4}{m_\rho^2}\bar\psi\psi\bar\psi\psi+\dots\\\no
&=&m_{ij}\bar\psi_i\psi_j\left(1+\Delta_{ij}\frac{\bar\chi}{f}+\dots\right)+
\frac{\lambda^4}{m_\rho^2}\bar\psi\psi\bar\psi\psi+\dots \ea The
appearance of the power of the dilaton follows from the expansion
of the operator $\int d^4y{\cal O}_R^a(x){\cal O}_L^b(y)$ in the
process of CFT integration. Again, if the dimensions
$\Delta_{L,R}$ were not to depend on the family label $a,b$, the
SILD coupling would be flavor diagonal
$\Delta_{ij}=\Delta_R+\Delta_L-4$. For a derivation of
eq.~(\ref{2}) in the Randall-Sundrum model see~\cite{Toharia}.

If the CFT is flavor anarchic we expect $\Delta_{ij}$ in~(\ref{1})
and (\ref{2}) to be a generic matrix with no hierarchical
relations between the components. We will assume this is the case
in the following. The leading FCNC effects can be thus described
by the following lagrangian: \ba\label{FCNC} \bar \psi_i\psi_j
\left[m_i\delta_{ij}\left(1+b\frac{\bar\chi}{f}\right)+\sqrt{m_im_j}b_{ij}\frac{\bar\chi}{f}\right]+C_{ijkl}\bar\psi_i\psi_j\bar\psi_k\psi_l,
\ea where the factor $\sqrt{m_im_j}$ in front of the flavor
violating term characterizes both classes of models illustrated above.

The flavor mixing term $b_{ij}$ is proportional to the anomalous
dimensions of the CFT operators, and it is expected to be an
$O(1)$ parameter in a strongly coupled dynamics. This parameter is
severely constrained by FCNC bounds. Tree level exchanges of the
dilaton lead to 4 fermions interactions with coefficients
$C\sim(mb^{ij}/f)^2/m_\chi^2$. The strongest bound for a generic
left-right mixing and CP violating operator comes from $K\bar K$
mixing. The UTFit analysis~\cite{UTFit} reports the bound
$Im(C)\lesssim(10^5$ TeV$)^{-2}$ (we are neglecting subleading
logarithmic running effects), which translates into \ba\label{b}
 |b_{ij}|\frac{v}{f}\lesssim10^{-2}\left(\frac{m_\chi}{100\, GeV}\right).
 \ea
If $v<f$, the bound~(\ref{b}) may be saturated for $b_{ij}=O(1)$
and a light dilaton. In this case the dominant decay channel is
the $\chi\rightarrow b\bar b$, and flavor violating processes may
have relatively large branching ratios $BR(\chi\rightarrow b\bar
s)\sim5\%$. In the regime $v\sim f$, the $b_{ij}$ should be
neglected compared to the flavor diagonal couplings. More
generally, the flavor violating coupling becomes
phenomenologically relevant at relatively large dilaton masses. In
the latter case, however, the dominant decay mode is expected to
be into massive vectors, as for a fundamental Higgs, and branching
ratios into fermions become much less accessible. One can estimate
flavor violating events to be down by an order $BR(\chi\rightarrow
t\bar c)\sim10^{-4}$ in this limit. Production channels like
$t\rightarrow\chi c$ may still play a role~\cite{Toharia}.

The 4-fermions contact terms in~(\ref{FCNC}) have coefficients of
the order $C_{ijkl}\sim \sqrt{y_iy_jy_ky_l}/m_\rho^2$, where
$m_i\sim y_iv$ is a \SM fermion mass. The phenomenological bounds
on these latter can be obtained from the bounds on the dilaton
couplings by noting the correspondence $m_\rho\sim
fm_\chi/(|b_{ij}|v)>10$ TeV. In a generic theory with heavy
composites around the scale $4\pi v$, this bound cannot be
satisfied and one should find a way to push the flavor violating
effects at higher scales. This can be done by adjusting the theory
in such a way that the only physics at the scale $g_\rho v\sim4\pi v$ is the strong Higgs sector, for which FCNC effects are negligible.
If our model can account for a
splitting between the CFT breaking scale and the \EW vacuum, the
4-fermion coupling $C_{ijkl}$ of~(\ref{FCNC}) would be suppressed
by a high scale $m_\rho=g_\rho f$ and the constraint~(\ref{b})
would translate into $f\sim$ TeV for a maximally strong dynamics
($g_\rho\sim 4\pi$).

To conclude, models of type~(\ref{class1}) are more adequate to
embed MFV, while those of type~(\ref{class2}) seem more
appropriate to explain the fermion hierarchy. From the above
qualitative estimates we see that $v/f\lesssim$ few$\times0.1$ in
generic models, while $v\sim f$ is compatible with FCNC
observations only if MFV is at work, in which case $b_{ij}=0$ by
definition.

\section{Higgs-dilaton mixing}

As far
as scale invariance is not explicitly broken, a hierarchical separation $v\ll f$ is unnatural in a CFT. Under this assumption, the dilaton is an exact \NG boson
and parameterizes a flat direction. Its effective lagrangian,
obtained after integrating out all the other fields, contains no
potential (see eq.~(\ref{self})): the only CFT invariant
candidate, $V=a\chi^4$, would imply an unbroken vacuum $\langle\chi\rangle=0$, that would contradict our original hypothesis $\langle\chi\rangle=f$.
The integration of light particles at the scale $v$ must therefore exactly
compensate the contribution of the heavy composites at the scale
$f$. Because by dimensional analysis the former give
$a_{light}\sim (v/f)^4$, the smaller the ratio $v/f$, the larger
the fine-tuning required to accommodate a vanishing potential.
This explains why, if the full \SM is embedded into a spontaneously broken CFT, the
natural cut-off scale for new physics cannot be far from the weak
scale. The same conclusion is found in the context of the \RS scenario if the \SM
is placed on the IR brane, where the cut-off suppressing higher
dimensional operators is naturally at the weak scale
irrespective of the mass of the Kaluza-Klein modes.

If the weak scale is somewhat smaller than $f$, and if the CFT is
not badly broken, it makes sense to consider an effective theory
for the strong EWSB sector of the SILD scenario and the dilaton
itself. We focus on a simplified theory in which the Higgs sector
is described by an interpolating Higgs doublet and for simplicity
we further assume that the anomalous dimensions can be algebraically summed\footnote{This can be made rigorous for appropriate local operators at leading
$1/N$ order and for a certain class of SUSY theories with
R-symmetry.}. Our
results are completely general, though.

Under our simplifying assumptions the most general action for the strongly coupled Higgs is constructed out of the conformally covariant
derivative~(\ref{der}): \ba
\left(D_\mu-\Delta\frac{\partial_\mu\chi}{\chi}\right)H, \ea where
$D_\mu=\partial_\mu+iA_\mu$ is the usual gauge covariant
derivative, and the potential \ba V(\chi,H)=\chi^4\hat
V\left(\frac{H^\dagger H}{\chi^{2\Delta}}\right). \ea Both kinetic
and potential terms induce a mixing between the Higgs after EWSB.
For the case of a dimension $\Delta=1$ Higgs, the kinetic mixing
coincides with the one induced in warped models by the non-minimal
gravitational coupling $\xi RH^\dagger H$ with
$\xi=1/6$~\cite{GRW}, see~\cite{CGK} for a detailed analysis. A
deviation from the conformal factor $\xi=1/6$ accounts for the
breaking of special conformal symmetries, and it amounts to
including an additional $(1-6\xi) H^\dagger D_\mu
H\partial^\mu\chi/\chi+h.c.$ operator in our 4D language.

In the conformal limit, the system can be easily diagonalized by
defining the scale-invariant field $\hat
H=H(f/\chi)^\Delta$. The mass eigenstates are then the physical component of the scale-invariant field, $\hat H^t=(0,v+\hat h)$, and the dilaton $\bar\chi=\chi-f$. Notice that, in order to be compatible with
the background solution $\chi=f$, $|H|^2=v^2$, the potential must
satisfy $\hat V=\hat V'=0$ on the vacuum. These conditions automatically ensure
that no mass term is generated for the dilaton. 

The leading
coupling between the dilaton and the physical
Higgs $\hat h$ is dictated by scale
invariance:
$$
\frac{m_\rho^2}{2}\hat
h^2\left(1+4\frac{\bar\chi}{f}+\dots\right),
$$
where $m_\rho^2=\hat V''v^2$. The mass of the physical composite Higgs $\hat h$ is
of order $m_\rho\lesssim4\pi v$ in a strong dynamics and its
fluctuation may be neglected at scales $E<m_\rho$. Because of the mixing $H-\chi$, all
the couplings of the gauge eigenvector $H$ induce vertices for
the SILD. In terms of the mass eigenstates, the electromagnetic singlet component of the doublet $H$ reads:  
\ba\label{h-d} h=(\hat h+v)\left(1+\Delta
\frac{\bar\chi}{f}+\frac{1}{2}\Delta(\Delta-1)\frac{\bar\chi^2}{f^2}+O(\bar\chi^3)\right).
\ea 
If we integrate out the heavy state $\hat h$, we see that the dilaton couples as a
Higgs up to a universal $\Delta v/f$ scaling and
modulo universality violations induced by the anomalous dimension
$\Delta-1$ of the strong Higgs. In particular, we see that the relation $a_1^2=a_2^2$ in~(\ref{gen}) characterizes the SILD scenario in the limit of small anomalous dimensions for the Higgs sector.

In principle, one may expect a large
dimension $\Delta$ to compensate the suppression $v/f$ so that $\Delta v/f\sim1$; however,
the dimension of the Higgs cannot be arbitrary large if we require
our model to be self-consistent. For example, assuming that the
top Yukawa coupling remains perturbative up to the naive
dimensional analysis scale $m_\rho\sim4\pi v$ requires
$\Delta<2$~\cite{CTC}.

Once sources of explicit symmetry breaking of the CFT are
taken into account, a non-zero dilaton mass $m$ is generated and
the decoupling procedure described above is no more valid. As far
as $m^2\ll m_\rho^2$ applies, we expect our results to be
accurate.

\paragraph{The dilaton as the Higgs:}

The quantity $v/f$ is an actual measure of the Higgs contribution
to the spontaneous CFT breaking, and parameterizes the amount of
mixing $h-\bar\chi$. This is explicit in~(\ref{h-d}) and it can be
equivalently derived as a general consequence of the algebra of
the generators, see eq.~(19) in~\cite{Fan}. For $v\sim f$ the
dilaton-Higgs mixing is maximal and in the extreme scenario $v=f$
the Higgs itself is the dilaton. 

The scenario in which the Higgs and the dilaton coincide naturally
account for a parametric separation between the Higgs mass and
the naive dimensional analysis scale $4\pi v$. However, both \EW precision parameters and
flavor measurements are potentially crucial tests for this class.
It is therefore of some interest to study the viability of this
scenario by considering explicit and computable models in which
this feature is realized. A tractable candidate may be found by considering the
Randall-Sundrum scenario.

In principle, in order to realize our program we
should be able to control the physics responsible for the IR brane generation.
This physics is related to operators with large dimensions --
$\Delta=O(N)$ -- which are dual to heavy stringy modes that we
cannot control on the 5D side~\cite{RZ}. Nevertheless, consistently with
our leading $1/N$ approximation we find that warped higgsless
theories~\cite{higgsless} behave as models in which the Higgs
field can be identified with the dilaton. To see this, we first
introduce the conformal coordinate $z=\log y$, in terms of which
the AdS geometry can be written as \ba\label{AdS}
ds^2=\frac{1}{z^2}\left(\eta_{\mu\nu}dx^\mu dx^\nu-dz^2\right),
\ea where the curvature has been normalized to 1 for simplicity.
In higgless models the Higgs sector can be idealized as a 5D field
$\phi$ infinitely peaked on the IR brane, which we assume to be
placed at some finite $z=z_{IR}$. Let us see how this translates in
the dual strong dynamics. 

Any bulk field $\phi$ on the gravity
side is believed to be dual to some operator ${\cal O}$ on the
gauge side. The 4D Fourier transform of a scalar field $\phi$ of
5D mass $m^2>0$ on the AdS geometry~(\ref{AdS}) is given by
$\phi(p,z)=c(p)z^2(J_\nu(pz)+\beta_p Y_\nu(pz))$, where
$\nu^2=4+m^2$ and $\beta_p$ is determined by the boundary
conditions. In the UV region $z\rightarrow0$ we have $\phi\rightarrow
Az^{\Delta}+\varphi_0z^{4-\Delta}$, where $\Delta=2+\nu$ represents the scaling dimension of the dual scalar operator
${\cal O}$ and corresponds to the positive
root of $m^2=\Delta(\Delta-4)$. The AdS/CFT correspondence instructs us to identify
the parameters $A$ and $\varphi_0$ as the vacuum expectation value
and the source of the operator ${\cal O}$,
respectively~\cite{Klebanov}.

By switching off the source
$\varphi_0$ and taking the limit of very large dimension $\Delta$ (i.e. large 5D mass, $m^2\rightarrow\infty$) we find $\phi(p,z)=c(p)z^2J_\nu(pz)\rightarrow
Az^{\nu+2}=A(p)z^\Delta$. Using the
identification introduced in~\cite{Klebanov} we finally write: \ba
\phi=\frac{\langle{\cal O}\rangle}{2\Delta-4}z^\Delta. \ea For a
translational invariant vacuum, no propagating mode
exists in the $m\rightarrow\infty$ limit and the bulk field reduces
to a non-dynamical background. Now, if the operator ${\cal O}$ acquires a
vacuum expectation value $\langle{\cal O}\rangle=cf^\Delta$, then
the corresponding 5D profile in the limit
$\Delta\rightarrow\infty$ can be approximated by a step function:
\ba
\phi\,\rightarrow\; \left\{ \begin{array}{ccc}  0\quad\;\,\texttt{for}\;\, zf<1, \\
\infty\quad\texttt{for}\;\,zf>1. \end{array}\right. \ea This is
exactly the physics at the origin of the IR brane in the RS model,
provided we identify $z_{IR}=1/f$~\cite{RZ}.

The argument can be generalized to the case in which a number of
operators ${\cal O}_i$ contribute to the spontaneous breaking of
the CFT by an amount $\langle{\cal O}_i\rangle=c_if^{\Delta_i}$.
As far as their dimension is very large $\Delta_i=O(N)$, the
excitations of these operators are not captured by our leading
$1/N$ approximations, and the information contained in the
coefficients $c_i$ is lost. In this limit the effective theory is
equivalent to having a single operator ${\cal O}_1$ with vacuum $f$. Any field
$\phi$ \textit{exactly} localized on the IR brane can be
interpreted as the dual description of ${\cal O}_1$. In particular, without
loss of generality, we can identify the Higgs as our candidate and conclude that
the above model is equivalent to a higgsless scenario of EWSB with $v=f$, as
anticipated. This model is characterized by \SM fields masses
of order $g_{SM}f$ and CFT composites at the scale $g_\rho f$. The
well known tension between perturbativity of the strong sector and
the fit with precision data -- in particular the flavor
constraints outlined in the previous section and the \EW precision
parameter -- is manifest. The question whether
subleading corrections in the $1/N$ expansion can alleviate this
tension while keeping a relatively light dilaton remains open.

\section{Discussion and conclusions}

If the forthcoming experiments discover a light and chargeless
CP-even scalar but no additional heavy states, it would be a
priority to understand whether the detected spin-0 particle is the
physical excitation of a light Higgs doublet, being it composite
or fundamental, or not. In order to answer this question we
focused on two scenarios of strong EWSB. The first includes the
fundamental Higgs doublet model and it is known as the SILH class~\cite{SILH}. The second scenario describes a broken
CFT with an emerging light dilaton~\cite{GGS}, the
strongly-interacting light dilaton (SILD) class.

The phenomenology
of the latter has been extensively studied in the context of the
\RS scenario, where it is generally accompanied by a light
composite Higgs (for a recent study see~\cite{csaki}). We find it
useful to attack the problem from the broader perspective of 4D
effective field theories.

Our primary aim was the identification of model-independent signals characterizing the dilaton with respect to the SILH scenario, questions regarding the compatibility of the models with the electro-weak precision parameters (in particular $\hat S$) were not considered. We used a phenomenological approach and characterized the models by using three parameters: the scale
at which the (approximate) symmetry of the strong dynamics is
spontaneously broken ($f$), the strength of the interactions among
the resonances ($g_\rho$), and the mass of the light scalar ($m$).
An explicit breaking of the global symmetries of the strong sector
is necessary to generate the mass $m$. The physically relevant aspect is how much
the \SM is involved in the breaking.

Given a competitive low energy phenomenology for the light
scalars (in particular, similar couplings to the standard model fermions), the two scenarios are in principle distinguishable at high
energies, as the heavy states are expected at parametrically
different scales, $g_{\rho}v\lesssim4\pi v$ for the SILD and
$g_{\rho}f\lesssim4\pi f$ for the SILH. In the low energy regime, however, a discrimination between a dilaton and a Higgs field, either
fundamental or composite, is possible 
in specific regions of the parameter space only as the result of combined fits and sufficiently high
statistics. A detailed estimate of the actual experimental significance of our
conclusions is a model-dependent issue and it has not been
addressed here.

At the LHC, the measurements of rates times branching ratios
($\sigma\times BR$) in many channels is expected to provide useful
pieces of information regarding the couplings of the light scalar
for a sufficiently high integrated luminosity. Under some reasonable assumptions on the new physics model\footnote{One of the assumptions that characterize the analysis of~\cite{Zepp} and~\cite{LPRZD} is that the couplings of the light scalar to two $W$'s or $Z$'s are bounded by above by those of the standard model Higgs. These assumptions are satisfied in both the dilaton scenario and the SILH (see~\cite{Low}), as far as our perturbative description holds.} and for an intermediate scalar mass $m$, it should be possible to probe relative deviations $\delta$
from the \SM Higgs couplings in the range $|\delta|>10\%$, see for instance~\cite{Zepp} and~\cite{LPRZD}. With these accuracies, if the SILD has an $f$ within $10\%$ or less of $v$, the LHC will not
be able to tell the difference between the two
scenarios~\cite{GGS}. A more promising environment is thus provided by
a linear collider like the ILC, at which a precision of the
percent level in the extraction of $\delta$ from $\sigma\times BR$
may be achieved.

Strong deviations from the SILH physics are likely to be observed
in the dilaton couplings to gluons or photons
$\chi\rightarrow2g,2\gamma$ mediated by the conformal
anomaly~\cite{GRW,GGS}. If the \SM gauge symmetry is an explicit
breaking of the CFT the situation is a bit more subtle because the
coefficients entering in such processes are very much
model-dependent. Nevertheless, we emphasize that observable departures in the production/decay
into massless vectors are expected to be relevant in theories
with a large number of fundamental constituents, which is quite a natural framework for obtaining a light dilaton.

A characteristic feature of the SILD -- or any model in which the
scalar is not part of a weak doublet -- would be the observation of
energy enhanced $O(4)$-violating processes in $V_LV_L\rightarrow\chi$ events (vector boson fusion as well as associated production) at high virtuality, for instance in the promising
channel $gg\rightarrow\chi\rightarrow
V_LV_L'\rightarrow4l,l^\pm\nu\nu$. To reach sufficiently high
energies, $p^2\gg m^2$, a hadron collider seems more adequate, but
the precision may not be sufficient due to QCD uncertainties and the large background.

We found that the dilaton generally mediates tree level flavor
violating processes in the effective theory. In contrast, a SILH
does not if the alignment mechanism described in~\cite{Contino}
applies. The flavor violating parameters are directly related to
the flavor non-universality of the CFT operators that couple to
the \SM fermions, and are present irrespective of whether the \SM
is or is not embedded into the CFT. Potentially testable branching
ratios for flavor violating processes like $\chi\rightarrow b\bar
s$ are natural features of a SILD. Other events characterizing the
dilaton model and involving \SM fermions were considered
in~\cite{Fan}. 

The CFT necessarily introduces 4-fermions contact
interactions in the low energy effective theory, as well, and the
current bounds on flavor violation can be translated into a bound
on the ratio $v/f$. Except for somewhat ad-hoc scenarios in which
the minimal flavor violation paradigm is at work, for which no
real explanation of the \SM fermion hierarchy is given, we found
that $v/f\lesssim$ few $0.1$ is a realistic estimate. Additional phenomenological bounds on this ratio come from LEP, and apply to a light dilaton $m<110$ GeV. These are discussed in~\cite{GGS}. 

In the preferable case $v<f$, a somewhat lighter Higgs sector compared to the scale
$m_\rho=g_\rho f$ is required. We analyzed the implications of
this assumption on the Higgs-dilaton mixing and studied the
decoupling of the physical excitations in the limit in which the
explicit CFT breaking source is negligible. It is intriguing to further consider the possibility of a large Higgs-dilaton mixing, $v\sim
f$, especially in light of a possible realization in a walking
dynamics context~\cite{Sannino}. We showed that the existing
warped higgsless models of EWSB~\cite{higgsless} are tractable incarnations of this
idea.

In summery, if no significant deviations in the magnitude of the \SM couplings
to the light scalar (especially in the gauge sector) are observed, no
$O(4)$-violating process is detected, as
well as no tree level FCNC exchange is measured, then it would be
fair to say that the observed scalar is a (composite) Higgs.

\acknowledgments
It is a pleasure to acknowledge Christophe
Grojean and Andreas Weiler for stimulating discussions, Riccardo Rattazzi for bringing this topic to my
attention, and Roberto Contino and Michael Graesser for
comments on the manuscript. This work has been partially supported by the Italian INFN and MIUR under the program "Fundamental
Constituents of the Universe" and the EU Network "UniverseNet"
(MRTN-CT-2006-035863), and by the U.S. Department of Energy at Los 
Alamos National Laboratory under Contract No. DE-AC52-06NA25396.

\appendix

\section{Parameterizations of the radion field}

In this Appendix we review the derivation of the radion profile in the Randall and Sundrum (RS) model and clarify its relations with the naive ansatz proposed by RS.

The RS model is constructed on a 5D space-time ranging from $y_+\leq y\leq y_-$ and possessing a $Z_2$ reflection symmetry around $y_+$ and $y_-$. Two branes are placed on the fixed points. The action of the model is $\int d^4x{\cal L}$, where
\ba\label{L}
{\cal L}&=&2\int_{y_+}^{y_-} dy \sqrt{-g}\left[-M^3R-V\right]+\sqrt{-g_+}\left[-V_+\right]+\sqrt{-g_-}\left[-V_-\right]+\dots
\ea
The dots stand for possible additional higher order terms and -- more importantly -- for the Gibbons-Hawking term, necessary to properly define the field theory on a space-time with boundaries. $g_{+,-}$ denote the determinants of the induced metrics at the points $y=y_{+,-}$ respectively, whereas the factor of $2$ in front of the integral over the extra dimension accounts for the orbifold symmetry. 

We focus on background solutions that preserve the 4D Poincare invariance. The most general line element with these properties reads
\ba\label{line} ds^2=e^{2A}\eta_{\mu\nu}dx^\mu
dx^\nu-dy^2, \ea where $A=A(y)$. 
The equations of motion for a line element of the form~(\ref{line}) reduce to the following independent equations:
\ba
A'^2=-\frac{V}{12M^3}\quad\quad A''=-\frac{V_+}{6M^3}\delta(y-y_+)-\frac{V_-}{6M^3}\delta(y-y_-),
\ea
where a prime means derivative with respect to the fifth coordinate $y$.
The RS background is defined by tuning the cosmological constants in such a way that $V=-12k^2M^3$, $V_+=-V_-=12kM^3$. Under these hypothesis our background solution is defined by $A=-k|y|$.

The gravitational perturbations around the RS background describe spin-2 and spin-0 fields, the spin-1 components being killed by the orbifold projection. The spin-2 fields form a tower of massive states with a single massless mode following from the preserved local 4D diffeomorphism invariance. The latter describes the fluctuations of the 4D background $\eta_{\mu\nu}$ and is thus properly identified with the 4D graviton. The spin-0 field, called radion, is a classically massless mode describing the fluctuations of the interbrane distance. Its mass gets non-trivial corrections if we go beyond Einstein gravity, namely the interbrane distance is generally stabilized by the introduction of higher dimensional operators. In terms of the dual variables~\cite{RZ}, the dilaton gets a mass once subleading ($1/N$) corrections are taken into account.

We are interested in studying the scalar degree of freedom. By an appropriate choice of gauge, a general line element can be put in the form
\ba\label{lineW}
ds^2=W^2\hat g_{\mu\nu}dx^\mu dx^\nu-Y^2dy^2,
\ea
where $W$ and $Y$ are arbitrary scalar functions of the 5D coordinates describing the dilaton fluctuations around the vacuum $\langle W\rangle=e^{-k|y|}$ and $\langle Y\rangle=1$. The traceless and transverse components of $\hat g_{\mu\nu}-\eta_{\mu\nu}$ describe the spin-2 modes. The standard procedure used to identify the dynamical modes consists in solving the linearized equations of motion for the perturbations $W-\langle W\rangle$, $Y-\langle Y\rangle$, and $\hat g_{\mu\nu}-\eta_{\mu\nu}$. The variation of the $\mu\nu$ part leads to the nontrivial condition
\ba\label{W1}
-\frac{1}{2}\frac{\partial_\mu\partial_\nu Y^2}{Y^2}-\frac{\partial_\mu\partial_\nu W^2}{W^2}+\hat g_{\mu\nu}(\dots)=0.
\ea
The $\hat g_{\mu\nu}$ term will not be reproduced here for brevity. It suffices to say that cancellation of this term, as well as of the $\mu5$ and $55$ components of the linearized equations, is ensured by imposing the constraint
\ba\label{WY}
W'=-kWY,
\ea
and by requiring that the linearized scalar 4D fields satisfy the massless Klein-Gordon equation. One can show that the two conditions~(\ref{W1}) and~(\ref{WY}) are sufficient to derive the radion profile~(\ref{rad}). The naive ansatz for the radion proposed by Randall and Sundrum, obtained by identifying the fifth coordinate $y$ with the radion, satisfies~(\ref{WY}) but not~(\ref{W1}), and thus it does not describe a dynamical mode.

The constraint~(\ref{WY}) will play the major role on our discussions to follow. To understand its physical meaning we focus on the (non-linear) physics of the radion and the 4D graviton. Without loss of generality we can assume $\hat g_{\mu\nu}=\hat g_{\mu\nu}(x)$ and write down the 5D lagrangian for these fields. Using the line element~(\ref{lineW}) we find:
\ba\label{formula}
-\sqrt{g}R=\sqrt{-\hat g}\left[-6\hat g^{\mu\nu}\partial_\mu\left(WY\right)\partial_\nu W-W^2Y \hat R+12\frac{W^2}{Y}W'^2+\left(\dots\right)'+\partial_\mu B^\mu\right],
\ea
where $\hat R$ is the 4D scalar curvature constructed out of $\hat g_{\mu\nu}$.
The last term in~(\ref{formula}) is a 4D boundary term, it has no perturbative implications and can be safely discarded in our treatment. The total derivative in $y$ exactly cancels the Hawking-Gibbons term, and thus can be ignored as well. The first and second terms describe at the quadratic level the radion and the graviton kinetic terms, respectively, and at the non-linear order the interactions of the two fields. The third term contains no 4D derivatives of the fields $W,Y$ and would represent a potential energy for the radion. This latter term must vanish once the contribution of the cosmological constants are included, and this gives~(\ref{WY}).

By solely imposing~(\ref{WY}) we thus find the form of the 4D effective action for the radion and the graviton:
\ba\label{eff}
{\cal L}_{eff}=\sqrt{-\hat g}\frac{M^3}{k}\left[6(\partial W)^2+W^2\hat R\right]^{y_-}_{y_+},
\ea
with $[f(y)]^{y_-}_{y_+}=f(y_-)-f(y_+)$. The action~(\ref{eff}) has been used by Rattazzi and Zaffaroni to spell out the dual interpretation of the physics of the IR brane~\cite{RZ}. These authors studied the effective theory~(\ref{eff}) in the absence of the UV brane regulator ($y_+\rightarrow-\infty$) and argued that the presence of the IR brane in the AdS background can be interpreted as a spontaneous breaking of the conformal symmetry of the strongly coupled 4D dual theory. In this limit the physical radion decouples from the UV (because of its composite nature) and can be identified with the dilaton $W(y_-)$.

From our derivation, it is clear that the relevant degree of freedom turns out to be $W$, rather than the actual dynamical 4D field $Q$ appearing in~(\ref{rad}). Any parametrization of the radion field having a small overlap with the UV brane ($\partial W(y_+)\ll\partial W(y_-)$) would be a suitable candidate to effectively describe the radion, up to hierarchically suppressed corrections. Furthermore, any parametrization of the radion would lead to the same non-derivative couplings for the 4D field. In particular, any parametrization would couple with IR strength to the IR brane, and would develop the same mass as the actual radion once explicit CFT breaking sources are added. This follows again from the condition~(\ref{WY}). Indeed, by the change of variables $y\rightarrow w(x^\mu,y)$, where
\ba
w=-\frac{1}{k}\log W
\ea
the line element~(\ref{line}) becomes
\ba
ds^2=e^{-2kw}\hat g_{\mu\nu}dx^\mu dx^\nu-dw^2+O\left(\partial_\mu w\right).
\ea
Hence, up to subleading derivative couplings and hierarchically suppressed corrections, the radion can be parametrized by any field $W$ satisfying~(\ref{WY}). Under these approximations, the physics of the naive ansatz and that of the dynamical mode $Q$ do coincide. Non-negligible corrections are expected to come into play only for small hierarchies, in which case the interpretation of the radion as a dilaton would fail.


 \end{document}